\documentclass[]{aa}
\usepackage{psfig}
\usepackage{txfonts}
\usepackage{graphicx}
\usepackage{amssymb}
\usepackage[below]{placeins}
\usepackage{natbib}
\bibpunct{(}{)}{;}{a}{}{,}

\def \arcmin{$^{\prime}$}

\begin{document}

\title{Complex X-ray morphology of Abell~3128:\\ A distant cluster behind a disturbed cluster}
\author{N. Werner\inst{1,2}
 \and E. Churazov\inst{2,3}
 \and A. Finoguenov\inst{4,5}
 \and M. Markevitch\inst{6}
 \and R. Burenin\inst{3}
 \and J. S. Kaastra\inst{1,7}
 \and H. B{\"o}hringer\inst{4}}

\offprints{N. Werner, email {\tt n.werner@sron.nl}}

\institute{     SRON Netherlands Institute for Space Research, Sorbonnelaan 2, NL - 3584 CA Utrecht, the Netherlands
\and		Max-Planck-Institute f{\"u}r Astrophysik, Karl-Schwarzschild-Strasse 1, 85749, Garching, Germany		
\and		Space Research Institute (IKI), Profsoyuznaya 84/32, Moscow, 117997, Russia
\and		Max-Planck-Institut f{\"u}r extraterrestrische Physik, 85748 Garching, Germany
\and 		University of Maryland Baltimore County, 1000 Hilltop circle, Baltimore, 21250, USA
\and            Harvard-Smithsonian Center for Astrophysics, 60 Garden St., Cambridge, MA 02138, USA
\and		Astronomical Institute, Utrecht University, P.O. Box 80000, NL - 3508 TA Utrecht, the Netherlands 
}

\date{Received, accepted }

\abstract{We present here the results of a detailed study of the X-ray properties of the cluster of galaxies Abell~3128 ($z=0.06$), based on the analysis of deep (100~ks) XMM-Newton data. The most obvious feature of the X-ray morphology of A3128 is the presence of two X-ray peaks separated by $\sim$12\arcmin. By detecting the redshifted Fe~K line, we find that the Northeast (NE) X-ray peak observed toward A3128 is a distant luminous cluster of galaxies at redshift $z=0.44$. Our subsequent optical spectroscopic observation of a distant radio bright galaxy in the centre of the NE X-ray peak with the Magellan telescope also revealed a redshift of $z=0.44$, confirming the association of the galaxy with the cluster seen in X-rays.
We detect a gravitational arc around the galaxy.
The properties of this galaxy indicate that it is the cD galaxy of the cluster in the background. The properties of the Southwest X-ray peak suggest that it is the core of a group merging with A3128 along our line of sight. Based on 2D maps of thermodynamic properties of the intra-cluster medium determined after subtracting a model for the background cluster, we conclude that an enhanced surface brightness region at a distance of $\sim$2.8\arcmin\ from the centre of the galaxy distribution is the centre of the gravitational potential of the cluster A3128. The unrelaxed nature of A3128 can be attributed to its location in the high density environment of the Horologium-Reticulum supercluster.

}

\maketitle

\section{Introduction}
\label{intro}

The formation and growth of the largest structures in the Universe, clusters of
galaxies and superclusters, is a still ongoing process.
According to the standard cosmological scenario clusters of galaxies continuously form and grow 
through merging with groups and individual galaxies, which are falling in along filaments. 
In several cases clusters are found close to each other in the highest overdensity regions of the Universe: the superclusters. In the crowded environment of superclusters, we can often study processes related to the growth of structure, like mutual interaction between the member clusters and their accretion history.

The Horologium-Reticulum (H-R) supercluster ($z=0.06$), together with the better known Shapley Supercluster are the
most massive structures in the local universe within a distance of 300~Mpc, 
with an estimated total mass of  $10^{17}$~M$_{\odot}$. \citet{einasto2001} list 35 member clusters in H-R.

Abell 3128 is a rich, highly substructured cluster in the H-R supercluster.
It has been well studied using redshift and radio surveys, and by Chandra \citep[20~ks;][]{rose2002}. 
Based on $N$-body simulations, \citet{caldwell1997} proposed that in the past 
A3128 encountered a tidal interaction with the cluster A3125 (current separation $1^{\circ}$, corresponding to $\sim$6~Mpc).
Rosat and Chandra observations of A3128 show a complex, disturbed X-ray morphology. Embedded in a $\sim$20\arcmin\ diameter diffuse halo
there are two cores in A3128, separated by 12\arcmin. 
The Northeast (NE) and Southwest (SW) cores comprise $\sim$85\% of the X-ray emission
of A3128, the remainder being in a diffuse halo. 
While the more luminous SW core is centred on a compact group around a bright elliptical galaxy, the NE core does not coincide with any bright galaxy. The NE core is asymmetric and
elongated, with the peak of the emission slightly displaced to the Northeast. With
core radii of $\sim$30~kpc, both cores are unusually narrow as compared to
other non-cooling core clusters with similar temperatures ($\sim$3.5~keV). 

\citet{rose2002} proposed a model to explain the double-peaked nature of the X-ray emission. In position-position and position-redshift diagrams they identified an infalling group and a candidate for a post-passage tidally distended group (filament), which they propose as candidates for producing the complex structure seen in X-ray images of the intra-cluster medium (ICM).  
They explain the SW core as the still intact hot gas of an infalling group.
The NE core represents according to \citet{rose2002} the surviving ICM of a group of galaxies that has fallen supersonically (Mach number 6)
into the cluster A3128 along the main filament connecting A3128 and A3125. The infalling group has, according to this scenario, passed through the core of A3128 and the galaxies of the group, moving
ballistically through the cluster, are by now well ahead of the stripped gas. 
The high infall velocity is caused by the deep potential well of the H-R supercluster, which makes even an infall of a small group an energetic and interesting event in the life of a galaxy cluster. \citet{rose2002} also report a possible detection of a radio arc, which lies slightly to the Northeast of the peak emission in the NE core, where the bow shock is expected if the gas is moving supersonically.

However, the Fe abundance of the NE core found by \citet{rose2002} is low (0.13 solar) and barring conclusive redshift information its association with A3128 could not be confirmed. As an alternative scenario to their unified view of the merging events occurring in A3128/A3125, \citet{rose2002} invoke the possibility that the NE emission peak is associated with a background cluster and the detected radio emission is associated with a radio bright central galaxy in a background cluster. In order to obtain a conclusive measure for placing of the NE component, a deeper observation for a redshift measurement using X-ray emission lines was needed.

Here we present the results of a deep 100~ks observation of Abell~3128 with the European Photon Imaging Cameras (EPIC) on XMM-Newton. The observation allows us to measure the redshift of the NE core and to unambiguously demonstrate that the dominant fraction of its emission originates in a background cluster. We also present here data obtained by the Magellan telescope, that allow us to confirm the association of the X-ray emission with a background cluster surrounding a radio bright cD galaxy.
The large effective area of XMM-Newton combined with the deep exposure, provides us with sufficient statistics to map the 2D distribution of thermodynamic properties of the ICM and thus to study the merging history of the disturbed, complex cluster of galaxies A3128. The XMM-Newton Reflection Grating Spectrometer data have insufficient statistics to derive accurate spectral properties for the cluster. 

Throughout the paper we use $H_{0}=70$ km$\, $s$^{-1}\, $Mpc$^{-1}$, $\Omega_{M}=0.3$, $\Omega_{\Lambda}=0.7$, which imply a linear scale of 78~kpc\, arcmin$^{-1}$ at the  redshift of A3128 \citep[$z=0.06$, ][]{rose2002}.  
Unless specified otherwise, all errors are at the 68\% confidence level for one interesting parameter ($\Delta \chi^{2}=1$). The elemental abundances are given with respect to the proto-solar values of \citet{lodders2003}.

\section{Observations and data analysis}

\subsection{XMM-Newton data}

Abell~3128 was observed with XMM-Newton during two pointings on May 29--30 and between May 31 and June 1st 2006 (revolutions 1185 and 1186) for a total exposure time of $\sim$100~ks. The two pointings were centred on the NE (72~ks) and on the SW core (32~ks), respectively. The EPIC/MOS detectors were operated in the full frame mode, while for the EPIC/pn detector the extended full frame window mode was employed. The observations were performed using the thin filter. The calibrated event files were obtained using the 7.0.0 version of the XMM-Newton Science Analysis System (SAS). For EPIC/MOS we keep only the single, double, triple, and quadruple pixel events (PATTERN$\leq$12), while for EPIC/pn, we make use of singe and double events (PATTERN$\leq$4). The spectral redistribution and ancillary response files are created with the SAS tasks {\texttt{rmfgen}} and {\texttt{arfgen}} separately for each camera and spectral extraction region that we analyze. 

Because of low number of counts in original bins, the extracted spectra are rebinned into bins with a minimum of 30 counts per bin. Our rebinned spectrum has bins of size at least 1/4 times the FWHM of the instrument. We fit the MOS1, MOS2, and pn spectra from both pointings simultaneously with the same model, with their relative normalizations left as free parameters. Our spectral analysis is restricted to the 0.4--7.0 keV band. In the spectral analysis we remove all bright point sources with a flux higher than $4.8\times10^{-14}$~ergs~s$^{-1}$~cm$^{-2}$.  

For the spectral analysis we use the SPEX package \citep{kaastra1996}. We fix the Galactic absorption in our model to the value deduced from the \ion{H}{i} data \citep[$N_{\mathrm{H}}=1.47\times10^{20}$~cm$^{-2}$, ][]{dickey1990}. We fit the cluster spectra with a plasma model in collisionally ionized equilibrium (MEKAL). The free parameters in the MEKAL model are the normalisation, temperature, and the Fe abundance. Since for elements other than Fe we can not obtain accurate abundance values, we fix them to 0.4 solar in the model. 

\subsection{Background modeling}

\subsubsection{Particle and instrumental background}

To clean the data from the soft proton induced events, we extract light curves for each detector separately in the 10--12~keV energy band where the cluster emission is negligible and the detected emission is dominated by the particle induced events. Since flares having a particularly soft spectrum may be missed when only the high energy part is studied, we also extract lightcurves in the 0.3--2.0~keV band. A visual inspection of the light curves reveals that the observation is not badly affected by the soft proton flares. After excluding the time periods when the count rate in the two considered energy ranges deviates from the mean by more than 3$\sigma$, we are left for the first pointing with 71~ks and 67~ks, and for the second pointing with 31~ks and 25~ks for EPIC/MOS and EPIC/pn, respectively. 

We subtract the EPIC instrumental background using closed-filter observations. For EPIC/MOS we use a closed filter observation (obs. ID 0150390101) with an exposure time of 200~ks. For EPIC/pn we use a closed filter observation (obs. ID 0106660401) with an exposure time of 120~ks. The instrumental background consists of fluorescence line emission, intrinsic instrumental noise, and particle induced noise caused by high-energy cosmic rays which are able to reach the detector even when the filter wheel is in closed position. The instrumental background varies from observation to observation. In order to scale the closed filter observations to the instrumental background during the source observation, we use the events registered outside the field of view (out-of-FOV) of the EPIC detectors, outside a radius of 15.4\arcmin\ from the FOV centre. 
Separately for each instrument, we scale the closed filter observation by the ratio of the 8--12~keV out-of-FOV count rate in our cluster observation and in the closed filter observation.

\subsubsection{The cosmic X-ray background}

\begin{table}
\begin{center}
\caption{The CXB components used in the fitting of the A3128 spectra. The power-law photon index is $\Gamma=1.41$. The unabsorbed fluxes are determined in the 0.3--10.0~keV band. 
\label{tab:back}}
\begin{tabular}{l|cccc}
\hline
\hline
Comp.			   &	k$T$   		& Flux     					\\
			   &  (keV)    		& (10$^{-12}$ erg cm$^{-2}$ s$^{-1}$ deg$^{-2}$) \\
\hline
LHB/SDC 		   & $0.08$  		& $2.5\pm0.6$ 	\\
HDC 	 		   & $0.25\pm0.02$      & $4.0\pm0.6$ 	\\  
EPL 			   &  	       		& $22.0\pm1.9$ 	\\

\hline
\end{tabular}
\label{tab:cxb}
\end{center}
\end{table}

We correct for the Cosmic X-ray Background (CXB) during spectral fitting.
\citet{kuntz2000} distinguish 4 different background/foreground components: the extragalactic power-law (EPL), the local hot bubble (LHB), the soft distant component (SDC), and the hard distant component (HDC). The EPL component is the integrated emission of faint discrete sources, mainly distant Active Galactic Nuclei (AGNs).
The LHB is a local supernova remnant, in which our Solar System resides. It produces virtually unabsorbed emission at a temperature of $\sim$10$^{6}$~K. The soft and hard distant components originate at larger distances. They might be identified with the Galactic halo, Galactic corona or the Local group emission and are absorbed by almost the full Galactic column density. Using the spectral band above 0.4 keV we can not reliably distinguish the SDC emission from the LHB component. Therefore, at temperatures below 0.1~keV we only consider the contribution of one thermal component. 

The large field of view, with regions where the contribution of cluster emission is small, enables us to determine the local properties of the background emission. 
We fit the spectra extracted from a region where the contribution of cluster emission is small and the emission is dominated by the X-ray background. 
We model the EPL emission with a power-law with a photon index of 1.41 \citep{deluca2004}. We model the soft foreground emission (LHB/SDC and HDC) by 2 collisionally ionized plasma models (MEKAL). Since the fitted spectral band does not allow us to accurately constrain the LHB/SDC temperature, we fix its value to 0.08~keV \citep[based on ][]{kuntz2000}. We leave the HDC temperature as a free parameter in the fit. To account for the remaining cluster emission in the extraction region, the contribution of which to the total flux is $\sim$30\%, we use an additional thermal model. We find that both the temperature and the flux of the emission attributed to the HDC ($0.25$~keV and $4.0\times10^{-12}$~erg~s$^{-1}$~cm$^{-2}$~deg$^{-2}$, respectively) are higher than the values determined based on the results published by \citet{kuntz2000} using the ROSAT All Sky Survey data (0.127~keV and $2.44\times10^{-12}$~erg~s$^{-1}$~cm$^{-2}$~deg$^{-2}$, respectively). This difference might be due to the contribution of the intra-supercluster medium. The adopted values of the temperatures and of the 0.3--10.0~keV fluxes of the background components are given in Table~\ref{tab:back}. In our subsequent spectral fits we fix the parameters of the background components to these adopted values.

\subsection{The X-ray images}
\label{image}

The spatial resolution of XMM-Newton allows a robust image reconstruction down to 8\arcmin\arcmin\ scales. However, due to the large wings of the point spread function (PSF), there is a significant contamination  from the emission on scales of 8\arcmin\arcmin\ or smaller to  large spatial scales upto several arcminutes.  For the purpose of cluster survey work (Finoguenov et al. in prep.), we have developed a procedure, which performs
an image restoration using  a symmetric model for the XMM PSF and  the calibration
of  \citet{ghizzardi2001}. The flux on small scales is estimated performing the scale-wise wavelet analysis, as described in \citet{vikhlinin1998}. Then the estimated flux is used to subtract the PSF model prediction on large scales and increase accordingly the flux on the small scales. The small scales in this procedure are a sum of 8\arcmin\arcmin\ and 16\arcmin\arcmin\ scales, which allows to avoid the variation in the PSF shape with off-axis angle, as described in \citet[][COSMOS special issue paper]{finoguenov2006}.

In Fig.1 we display both the large scales of the emission starting at the 32$^{\prime\prime}$ scale together with smaller scales (8$^{\prime\prime}$ and 16$^{\prime\prime}$).

\section{Results}

\begin{figure*}
\centering
\includegraphics[width=15cm]{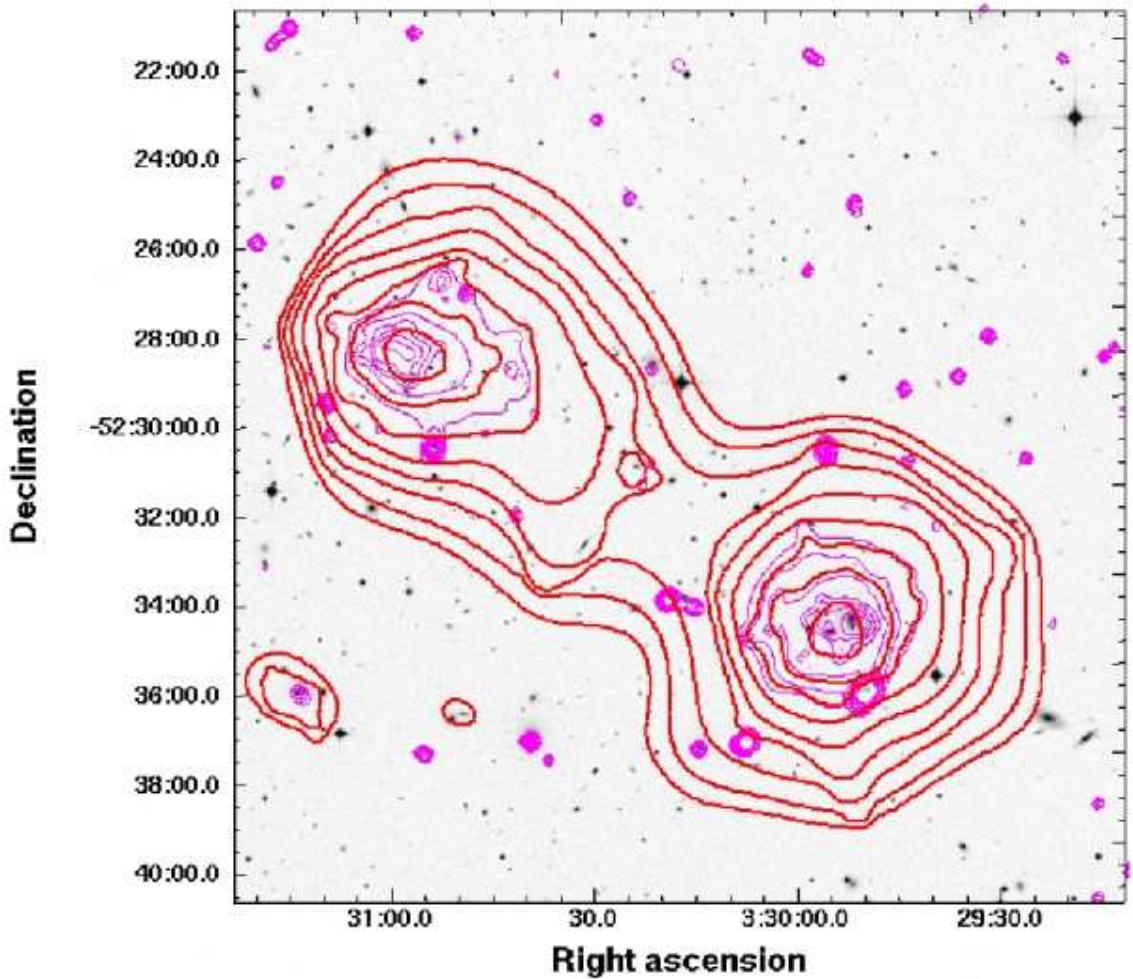}
\caption{X-ray contours superimposed on the Digitalized Sky Survey image of A3128. North is up, West is to the right. The thick contours highlight the large scale features in the X-ray morphology, the thin contours highlight the features at small spatial scales. The contour spacing is arbitrary. }
\label{xrayopt}
\end{figure*}

The most obvious feature of the X-ray morphology of A3128 is the presence of two X-ray peaks separated by $\sim$12\arcmin. In Fig.~\ref{xrayopt} we show the X-ray contours superimposed on the Digitalized Sky Survey image of the cluster. The thick contours highlight the large scale (32$^{\prime\prime}$) features in X-ray morphology, the thin contours highlight the features on small spatial scales (8$^{\prime\prime}$ and 16$^{\prime\prime}$).

While the SW peak of the X-ray intensity coincides with a bright elliptical galaxy (ENACS~75) surrounded by a compact group of galaxies, the NE core does not coincide with any bright galaxy. On the small spatial scales the NE peak has a strongly elongated morphology and on larger scales it seems to have a tail toward the SW direction. The thick contours also reveal an association of the hot gas with two small groups of galaxies to the East of the main cluster contours.

\subsection{Properties of the X-ray peaks}

\subsubsection{The global properties}
\label{globprop}

\begin{figure*}[tb]
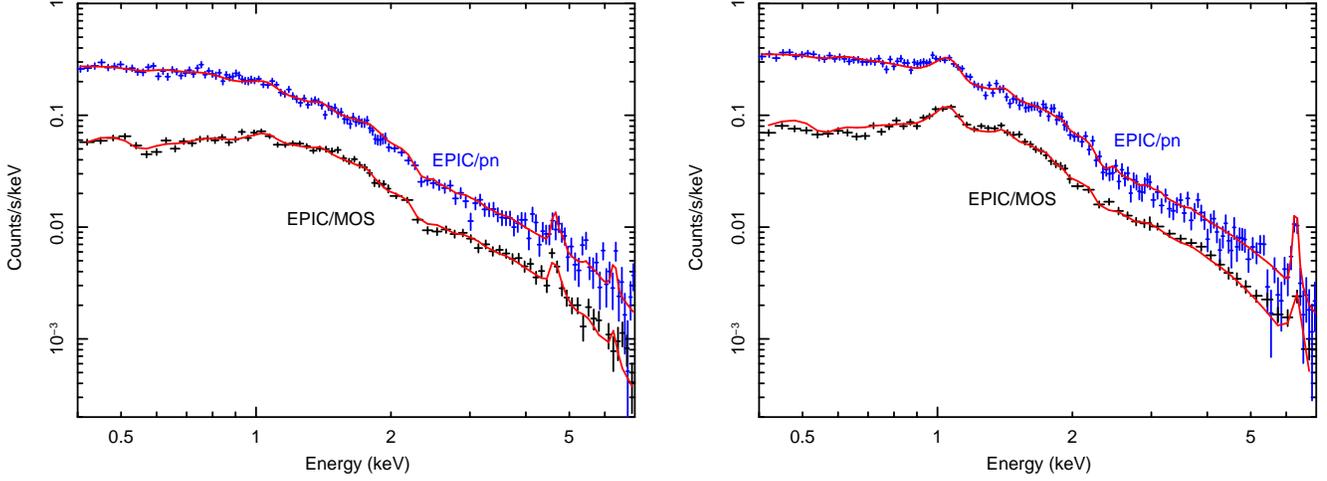

\begin{minipage}{0.5\textwidth}
\includegraphics[width=0.7\textwidth,clip=t,angle=270.]{NE.ps}
\end{minipage}
\begin{minipage}{0.5\textwidth}
\includegraphics[width=0.7\textwidth,clip=t,angle=270.]{SW.ps}
\end{minipage}\\
\caption{The spectrum of the NE (left panel) and SW (right panel) X-ray peak extracted from circular region with a radius of 1.5\arcmin. The continuous line represents the best fit model. The EPIC/pn and the coadded EPIC/MOS spectra are indicated. The position of the Fe~K line in the spectrum from the NE peak shows that the emitting cluster is at a redshift of $z=0.44$.} 
\label{spectra}
\end{figure*}

\begin{table}
\caption{Fit results for the spectra extracted from the SW X-ray peak, from the apparent ``tail'' of the NE peak toward SW, and from the NE X-ray peak. The spectra extracted from the SW peak and from the ``tail'' region were fitted with a thermal model  with the $N_{\mathrm{H}}$ fixed to the Galactic value \citep[$N_{\mathrm{H}}=1.47\times10^{20}$~cm$^{-2}$, ][]{dickey1990}. The NE X-ray peak was fitted with two thermal models: one for the emission from A3128 in the foreground and one for the background cluster. In this fit the redshift, temperature, and the Fe abundance of the foreground cluster were fixed to the best fit values from the ``tail'' region. The reported best fit parameter of the NE peak describe the properties of the background cluster. The temperature is given in keV, the emission measure $Y$ is given in units of $10^{65}$~cm$^{-3}$ and the Fe abundance is given with respect to
the proto-solar values of \citet{lodders2003}.}
\begin{center}
\begin{tabular}{l|ccc}
\hline\hline
Par.              &  SW 	     & ``tail''        &  NE		   \\
\hline
$Y_{z=0.06}$      &  $9.72\pm0.10$   & $6.80\pm0.07$    & $3.4\pm0.4$	    \\
$Y_{z=0.44}$      &  -- 	     &  --	       & $303\pm34^\dagger$	    \\
k$T$              &  $3.36\pm0.05$     & $3.50\pm0.08$ & $5.14\pm0.15^\dagger$      \\
Fe                &  $0.69\pm0.04$   & $0.49\pm0.04$   & $0.47\pm0.08^\dagger$      \\
$z$	          &  $0.060\pm0.001$ & $0.058\pm0.003$  & $0.444\pm0.003^\dagger$     \\
\hline
$\chi^2$ / $\nu$  & 652/501	     & 435/398         & 943/800	    \\
\hline
\end{tabular} 
\end{center}
{\footnotesize{$^\dagger$ values determined for the background cluster.}}
\label{tabglob}
\end{table}

In order to study the global spectral properties of the two X-ray bright peaks, we extract their spectra from circular regions with a radius of 1.5\arcmin. In Fig.~\ref{spectra} we show the spectra with their best fit model. Previous Chandra data showed that while the SW X-ray core has a high Fe abundance, the Fe abundance of the NE core is very low, and therefore it is also not possible to confirm its association with the cluster A3128 using the redshift of the Fe~K line. Even using the deep XMM-Newton data we do not see a clear Fe~K emission line at the expected energy for the redshift of A3128. However, we detect in all EPIC detectors for both XMM-Newton pointings a line at $\sim$4.6~keV. This corresponds to the Fe~K line emission of a cluster at a redshift of $\sim$0.45. 

We also extract a spectrum from a circular region with a radius of 2\arcmin\ centred on the apparent tail of the NE X-ray peak directed to the SW ($\alpha=3^{\mathrm{h}}30^{\mathrm{m}}30^{\mathrm{s}}$, $\delta=-52^{\circ}31$\arcmin$0$\arcmin\arcmin). We immediately see an Fe~K line at $\sim$6.3~keV and we see no obvious line feature around 4.6~keV. We fit the spectrum extracted from the ``tail'' region and from the SW X-ray peak with a thermal model. The free parameters of the fit are the redshift, temperature, Fe abundance, and emission measure. The best fit parameters are shown in Table~\ref{tabglob}.

We fit the spectrum of the NE peak with a combination of two thermal plasma models: one for the emission of A3128 and the other for the emission of the background cluster. We fix the parameters of the model of the foreground cluster to the values determined for the tail region: redshift $z=0.06$, metallicity $Z=0.49$, and temperature k$T=3.5$~keV. The emission measure of the foreground cluster and the redshift, Fe abundance, temperature, and emission measure of the background cluster are free parameters in the spectral fit. The best fit redshift value for the background cluster is $z=0.444\pm0.003$. The linear scale at this redshift for the $\Lambda$CDM cosmology is 341~kpc\, arcmin$^{-1}$. The best fit temperature, Fe abundance, and emission measure of the background cluster are shown in the column ``NE'' of Table~\ref{tabglob}. Our spectral analysis indicates that $\sim$60\% of the X-ray emission in the extraction region is coming from the background cluster. The 0.3--10.0~keV luminosity of the background cluster within the radius of 1.5\arcmin\ (512~kpc) is $4.7\times10^{44}$ ergs~s$^{-1}$.

The best fit parameters of the background cluster depend on the assumptions about the emission of the foreground cluster. Therefore, the systematic uncertainties on the best fit parameters are larger than the quoted statistical uncertainties. For a lower assumed temperature for the foreground cluster, we fit a higher temperature for the background cluster. For example, if we assume that the temperature of the foreground cluster is 3.0~keV instead of 3.5~keV, the best fit temperature of the background cluster will be 5.8~keV instead of 5.1~keV. 

We note, that there is no evidence of emission from the background cluster in the ``tail'' region. From the spectral analysis we find that its contribution in this region is less than 1\%.

\subsubsection{Radial profiles}
\label{secprofiles}

\begin{table}
\caption{Radial profiles determined for the SW X-ray peak. The temperature is given in keV, the emission measure $Y$ is given in units of $10^{65}$~cm$^{-3}$, and the Fe abundance is given with respect to the proto-solar values of \citet{lodders2003}.} 
\begin{center}
\begin{tabular}{l|cccc}
\hline\hline
Par.	& 0.0\arcmin--0.5\arcmin\ & 0.5\arcmin--1.5\arcmin\ & 1.5\arcmin--2.5\arcmin\ & 2.5\arcmin--3.5\arcmin\  \\
\hline
$Y$  & $1.99\pm0.05$ & $7.91\pm0.09$ & $8.67\pm0.10$ & $9.57\pm0.12$ \\
k$T$ & $3.36\pm0.11$ & $3.36\pm0.06$ & $3.17\pm0.06$ & $3.38\pm0.07$ \\
Fe   & $0.87\pm0.10$ & $0.67\pm0.04$ & $0.48\pm0.06$ & $0.46\pm0.04$ \\
\hline
$\chi^2$ / $\nu$ & 269/229 & 1125/959 & 1183/1072 & 1273/1137 \\
\hline
\end{tabular}
\label{tabprofSW}
\end{center}
\end{table}

\begin{table}
\caption{Radial profiles determined for the NE X-ray peak. The metallicity and the temperature of the foreground cluster were fixed to 0.49 solar and 3.5~keV, respectively. 
The emission measure $Y$ is given in units of $10^{65}$~cm$^{-3}$ and the temperature is given in keV. The Fe abundance is given with respect to the proto-solar values of \citet{lodders2003}. The $f_{z=0.44} / f_{z=0.06}$ indicates the ratio of the fluxes from the background cluster and A3128 in the given extraction region. The superscript ``$f$'' indicates that the parameter value was fixed during the fitting.}
\begin{center}
\begin{tabular}{l|cccc}
\hline\hline
Par.	& 0.0\arcmin--0.5\arcmin\ & 0.5\arcmin--1.5\arcmin\ & 1.5\arcmin--2.5\arcmin\ & 2.5\arcmin--3.5\arcmin\  \\
\hline
$Y_{z=0.06}$  & $0.56\pm0.20$ & $2.7\pm0.4$  & $4.7\pm0.4$   & $5.6\pm0.3$ \\
$Y_{z=0.44}$  & $72\pm15$     & $178\pm27$   & $154\pm34$    & $<140$ \\
k$T$          & $5.0\pm0.3$   & $5.7\pm0.3$  & $3.5\pm0.3$   & $4^{f}$ \\
Fe            & $0.56\pm0.18$ & $0.49\pm0.09$& $0.21\pm0.11$  & $0.2^{f}$ \\
$\frac{f_{z=0.44}}{f_{z=0.06}}$ & 2.1 &  1.2  &  0.55 & 0.09 \\ 
\hline
$\chi^2$ / $\nu$&   260/266   & 827/749  & 986/841   &  1109/870 \\
\hline
\end{tabular}
\label{tabprofNE}
\end{center}
\end{table}

We determine the emission measure, temperature, and Fe abundance profiles for both X-ray peaks. We extract the spectra from circular annuli with outer radii of 0.5\arcmin, 1.5\arcmin, 2.5\arcmin, and 3.5\arcmin. 

In Table~\ref{tabprofSW} we show the best fit parameters determined for the SW X-ray peak. The temperature profile of this peak is flat and there is no evidence for a cool core. However, the Fe abundance has a strong peak in the centre of the X-ray emission, which coincides with an elliptical galaxy surrounded by a group of galaxies. The abundance peaks within the central 1.5\arcmin, which corresponds to $\sim$120~kpc. 

Fitting a beta model \citep{cavaliere1978} to the radial surface brightness distribution of the SW X-ray peak we find $\beta=0.3$ and a core radius of $r_{c}=30$~kpc. As already noted by \citet{rose2002} this core radius is about an order of magnitude lower than the more typical values of $\sim$250~kpc that are observed for non-cooling core clusters of galaxies. The $\beta$ value is also low compared to the typical value of $\sim$0.7. 
The small value of $\beta$ shows that the surface brightness distribution of the core is broader than that usually observed for clusters, which indicates that we might see the emission of a group superimposed on the diffuse cluster emission. 

Using the parameters of the $\beta$ model fit, and a global temperature value of k$T=3.4$~keV, we find that the mass enclosed within the radius of 120~kpc is $1.2\times10^{13}$ M$_{\odot}$. The total gas mass within the same volume, determined using a central electron density of $1.2\times10^{-2}$~cm$^{-3}$, is $\sim$$1.3\times10^{12}$~M$_{\odot}$.
Using the same parameters, the estimated total mass within a radius of 1~Mpc is $\approx$$1.1\times10^{14}$~M$_{\odot}$. 
A radius of 1~Mpc is approximately the virial radius of a 3.4~keV cluster \citep[e.g.][]{finoguenov2001}. 

The total Fe mass enclosed within the radius of 120~kpc, the region within which the Fe abundance peaks, is $\sim$$1.5\times10^9$~M$_{\odot}$. Assuming a flat Fe abundance distribution in A3128 of 0.45 solar, the excess Fe mass in the SW X-ray peak is $\approx$5$\times10^8$~M$_{\odot}$. 

The radial profiles determined for the background cluster seen as the NE X-ray surface brightness peak are shown in Table~\ref{tabprofNE}. For the temperature and Fe abundance of the foreground cluster we assume 3.5~keV and 0.49 solar, respectively. The free parameters in the fit are the temperature and the Fe abundance of the background cluster, and the emission measures of both clusters. We see an indication of a temperature drop in the inner 170~kpc, then the temperature peaks, and outside 500~kpc it drops again. The spectra indicate that the Fe abundance peaks in the cluster core.
In Table~\ref{tabprofNE} we also show for each annulus the ratio of the flux of the background cluster with respect to the flux of the foreground cluster.
In the innermost extraction region the emission of the background cluster clearly dominates, between 0.5\arcmin--1.5\arcmin\ the contributions of both clusters are similar. Between 1.5\arcmin--2.5\arcmin\ the emission of the foreground cluster dominates, and between 2.5\arcmin--3.5\arcmin\ we can only determine an upper limit for the emission of the background cluster. The total 0.3--10.0~keV luminosity of the $z=0.44$ cluster within the radius of 3.5\arcmin\ is $6.9\times10^{44}$~ergs~s$^{-1}$.  

To obtain a rough estimate of the total mass of the background cluster, we fit a beta model to the radial profile of its best fit emission measure per arcminute. 
The best-fit parameter values are $r_c=157$~kpc and $\beta=0.41$. 
Assuming a global temperature of 5.14~keV (the best fit global temperature from Sect. \ref{globprop}) we obtain a total mass within a radius of 1.5~Mpc \citep[estimated $r_{500}$ of a 5.1~keV cluster, e.g. see][]{finoguenov2001} of $3.4\times10^{14}$~M$_{\odot}$. We note that due to the uncertainties associated with the subtraction of the foreground cluster, and because the core of the background cluster is far from being relaxed and it is not single-temperature, this derived mass is only a very rough estimate.

\subsection{Optical observation of the background cluster}
\label{sec:optical}

\begin{figure}
  \centering
  \includegraphics[height=5.7cm]{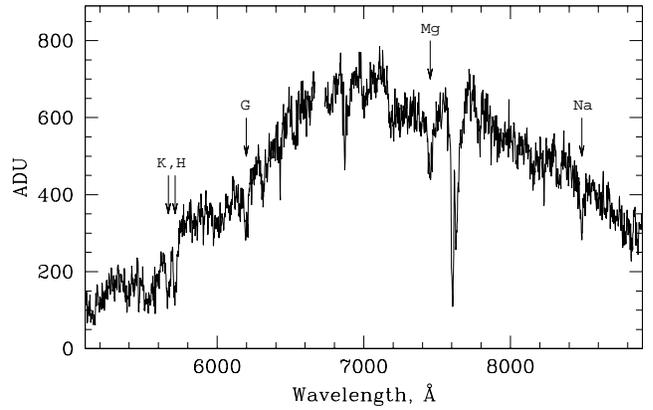}
  \caption{Spectrum of the galaxy associated with the radio source SUMSS~J033057-522811 in the centre of the NE X-ray peak. On the y-axis we plot arbitrary count units (ADU). Lines, used for redshift determination, are shown with arrows.}
  \label{fig:sp}
\end{figure}

\begin{figure}
  \centering
  \includegraphics[height=5.7cm]{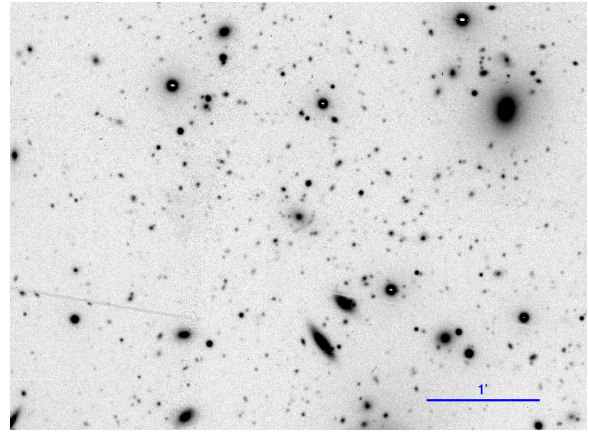}
  \caption{Combined R-band image of the galaxy associated with the
radio source SUMSS~J033057-522811 in the centre of the NE X-ray peak. The total exposure time is 360~s. Note the arc to the Southwest of the galaxy. The scale of 1\arcmin\ is indicated on the image. }
  \label{fig:r}
\end{figure}

Optical spectroscopic observations of the galaxy associated with the
radio source SUMSS~J033057-522811 \citep{mauch2003} were done with the
6.5~m Magellan I Baade telescope, using the Inamori Magellan Areal
Camera and Spectrograph (IMACS). We used the short f/2 camera with a
$27^\prime\times27^\prime$ FOV and $0.2^{\prime\prime}$ pixel
scale. In spectroscopic mode this setup with 300 l/mm grism and
$1.2^{\prime\prime}$ slit gives a spectral resolution of $\approx6$~\AA.

The spectrum obtained is shown in Fig.~\ref{fig:sp}. It is typical for
an early type galaxy and does not show detectable emission
lines. Using the standard set of absorption lines, marked in
Fig.~\ref{fig:sp} with arrows, the redshift $z=0.43961\pm0.00014$ was
derived. This value is in a very good agreement with the X-ray
determined redshift thus providing immediate confirmation that the
observed X-ray emission is due to the emission of a distant cluster.

During the slit alignment procedure we also obtained a few direct R-band
images of this field. The combined image is shown in
Fig.~\ref{fig:r}. The field was photometrically calibrated using observations of Landolt standard
stars \citep{landolt1992}. We measured the magnitude of the
galaxy to be $m_{\mathrm{R}} = 18.78$ (in 4\arcmin\arcmin aperture), which is approximately
what is expected for a brightest cluster cD galaxy at this redshift \citep[e.g.][]{vikhlinin1998}. The absolute magnitude of the
galaxy in R was estimated to be -23.35 \citep[adopting K and evolution
corrections from][]{poggianti1997}. For comparison the absolute
magnitudes of M~87 and Cygnus A are approximately -23.2 and -23.3
respectively - close to the absolute magnitude of SUMSS~J033057-522811.

The gravitational lensing arc is clearly seen around this galaxy
(Fig.~\ref{fig:r}) at a radial distance of $\sim 6.2''$. The enclosed
mass within this radius can be estimated \citep[e.g.][]{narayan1996} as
\begin{eqnarray}
M=1.1 \times 10^{14} M_\odot \left( \frac{\theta}{30''}\right)^2 \left ( \frac{D}{1~{\rm Gpc}}\right) 
\end{eqnarray} 
where $\theta=6.17''$ is the arc radius and $D=D_{\mathrm{d}} D_{\mathrm{ds}} / D_{\mathrm{s}}$
is the combination of (angular diameter) distances from the observer
to the lens $D_{\mathrm{d}}$ and to the source $D_{\mathrm{s}}$, and from the lens to the source
$D_{\mathrm{ds}}$. In our case only $D_{\mathrm{d}}=1.17~{\rm Gpc}$ is known. 
For $D=1$ Gpc we obtain an upper limit to the total mass of $M < 5\times10^{12}$~M$_\odot$ within $\sim$35~kpc of SUMSS~J033057-522811. 
This value is about a factor of 2 higher than the total mass (within a similar distance) of M~87 \citep{matsushita2002}, derived using
X-ray data.

One can therefore conclude that optical data strongly suggest that
SUMSS~J033057-522811 is a massive elliptical galaxy with the
parameters characteristic for most bright cD galaxies in the local
Universe. 

In the Magellan image we also see an excess of the surface density of faint
galaxies near this cD galaxy, most of which are probably members of the distant cluster.

\subsection{2D maps of thermodynamic properties}

\begin{table}[tb]
\caption{The best fit values determined for the X-ray bright low entropy region and for the region at the centre of the galaxy distribution. 
The emission measure $Y_S$ per square arcminute is given in units of $10^{64}$~cm$^{-3}$~arcmin$^{-2}$ and the temperatures are in keV. The Fe abundance is given with respect to the proto-solar values of \citet{lodders2003}. The entropy $s$ is given in keV~cm$^{2}$ and the pressure $P$ is given in 10$^{-12}$~dyne~cm$^{-2}$.}
\begin{center}
\begin{tabular}{l|cc}
\hline\hline
Par.	& X-ray bright reg. & centre of gal. dist. \\
\hline
$Y_S$  	        & $7.54\pm0.08$  & $4.03\pm0.06$ \\
k$T$            & $3.30\pm0.08$ & $3.79\pm0.13$\\
Fe              & $0.43\pm0.04$ & $0.53\pm0.07$ \\
$s$		&   223         &  335          \\
$P$		&   9.5         &  7.3           \\
\hline
$\chi^2$ / $\nu$&  563/569 & 448/460 \\
\hline
\end{tabular}
\label{lowhighent}
\end{center}
\end{table}

\begin{figure*}
\begin{minipage}{0.5\textwidth}
\includegraphics[width=0.85\textwidth,clip=t,angle=0.]{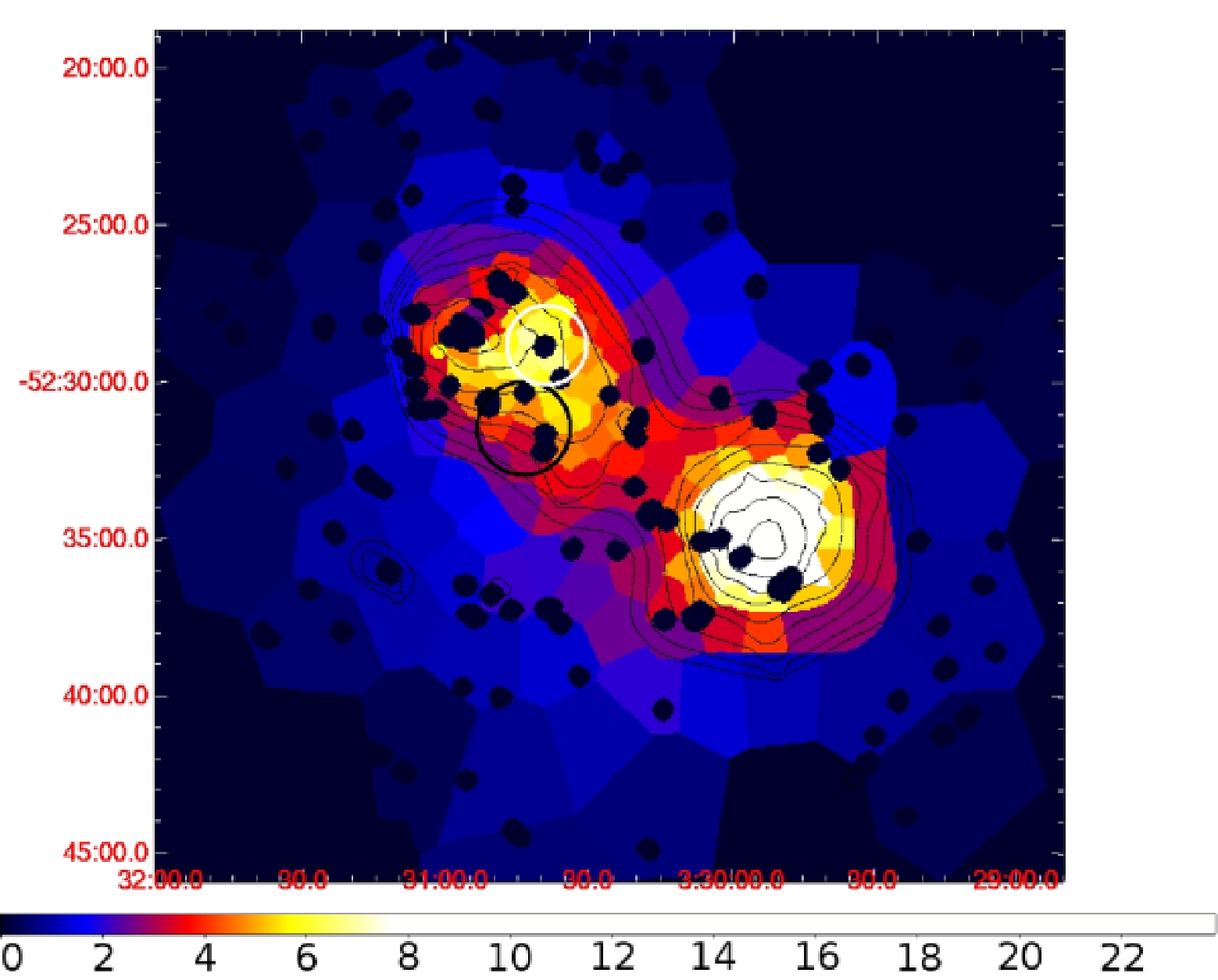}
\end{minipage}
\begin{minipage}{0.5\textwidth}
\includegraphics[width=0.85\textwidth,clip=t,angle=0.]{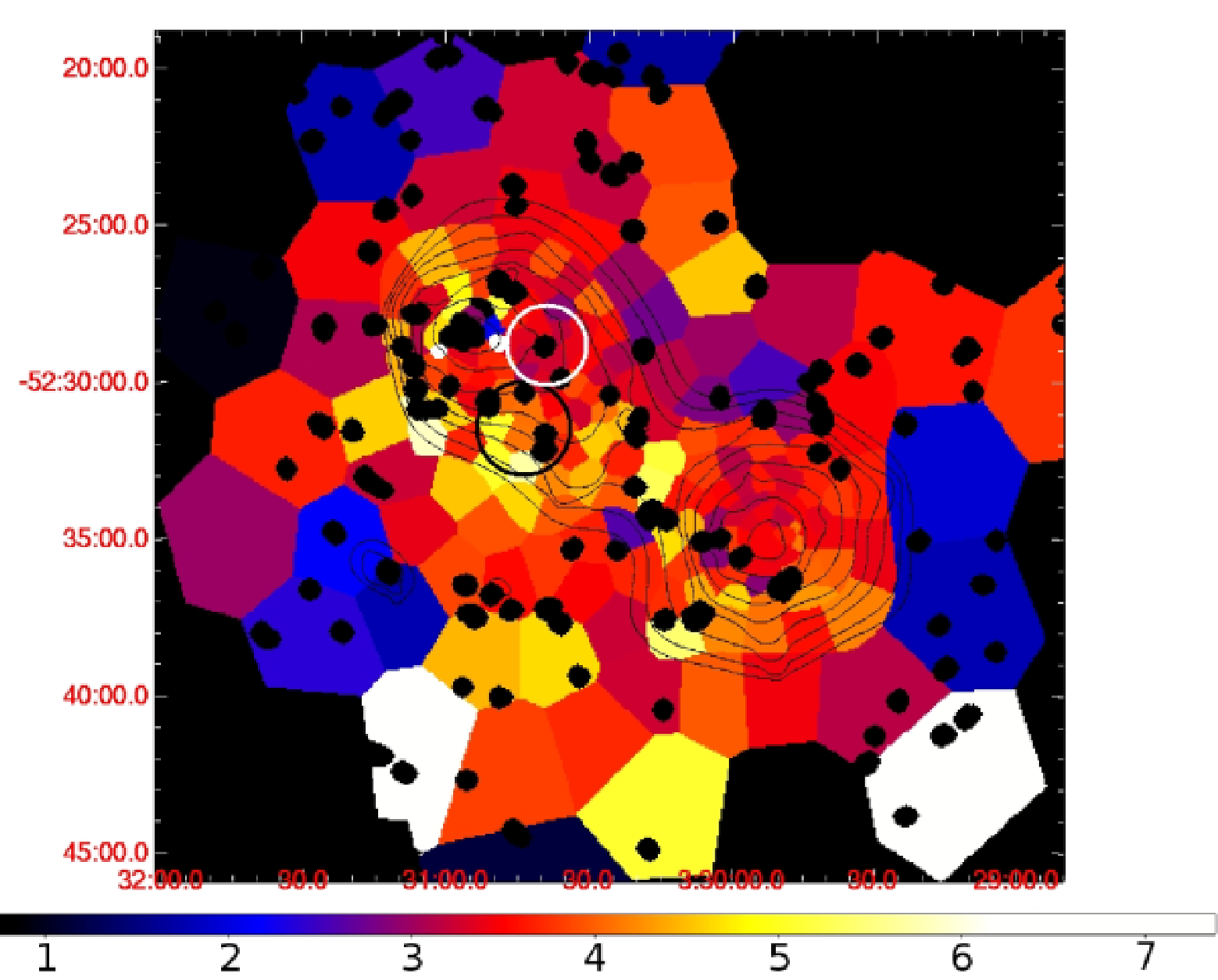}
\end{minipage}\\
\vspace{2mm}

\begin{minipage}{0.5\textwidth}
\includegraphics[width=0.85\textwidth,clip=t,angle=0.]{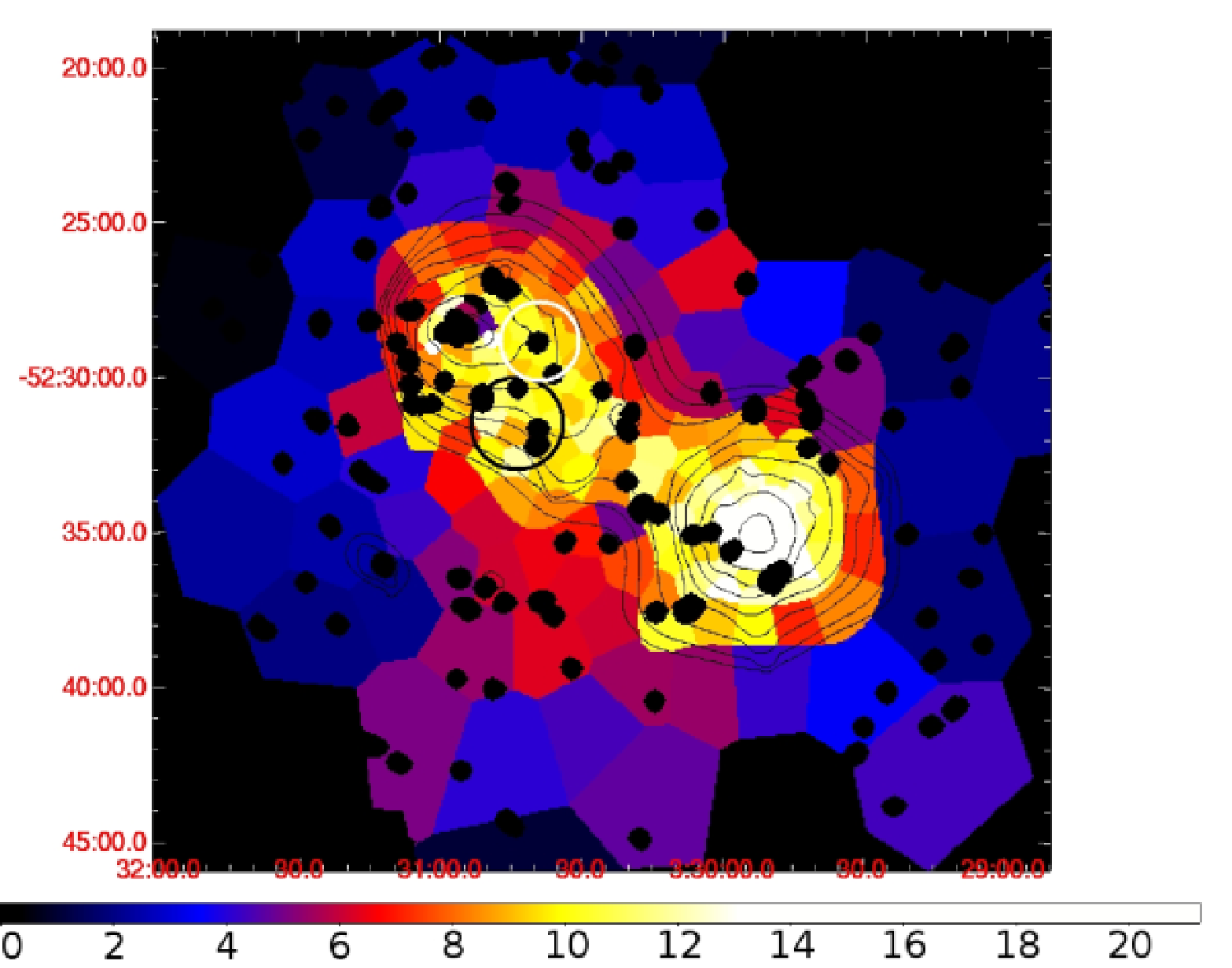}
\end{minipage}
\begin{minipage}{0.5\textwidth}
\includegraphics[width=0.85\textwidth,clip=t,angle=0.]{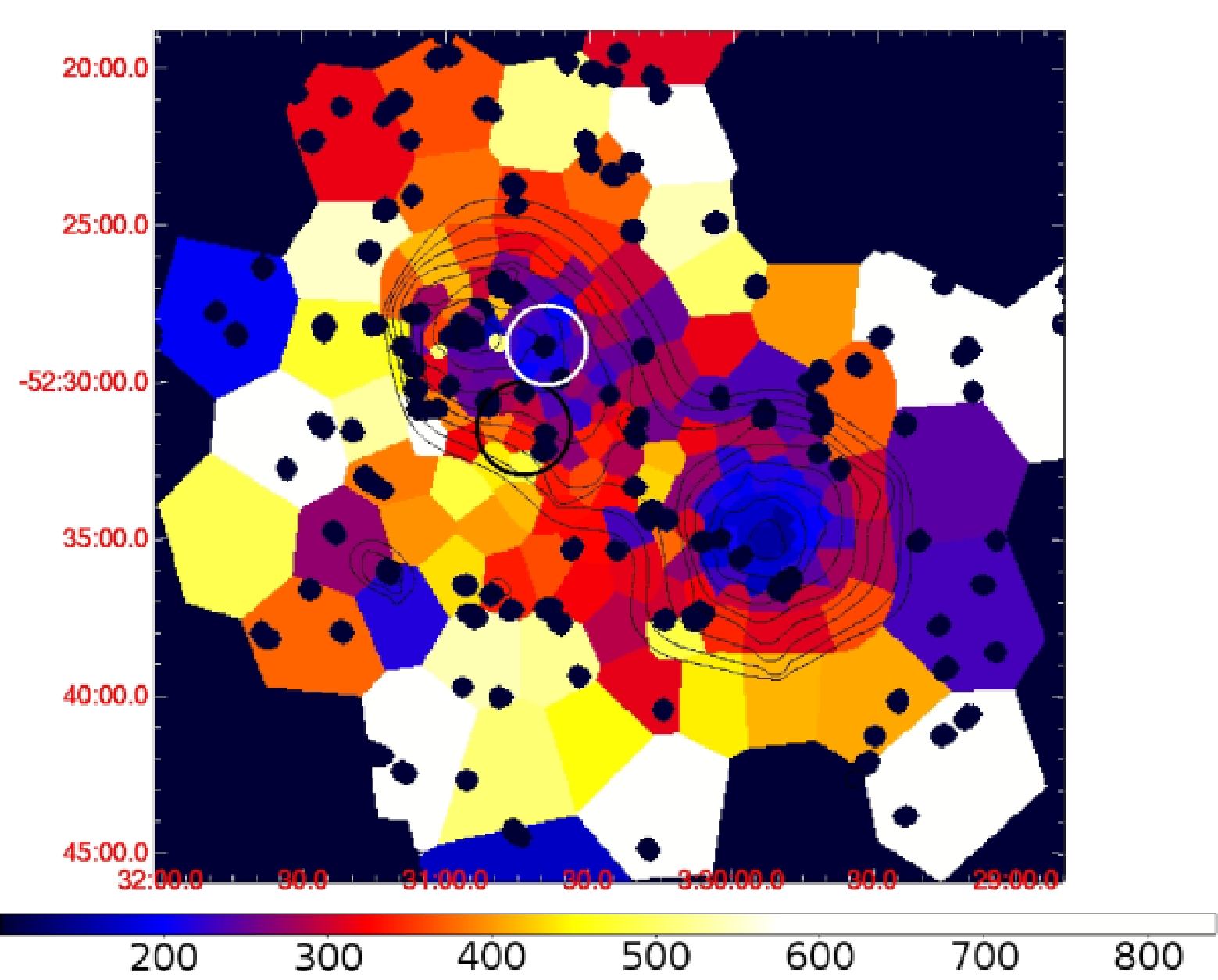}
\end{minipage}
\caption{{\it{Top left: }}The projected emission measure per arcminute, in units of $10^{64}$~cm$^{-3}$~arcmin$^{-2}$, integrated along the line of sight, after the emission of the background cluster was subtracted. {\it{Top right: }}Temperature map derived from spectral fitting in units of keV. {\it{Bottom left: }}Map of the pressure derived from the best fit temperature and emission measure, assuming a length scale along our line of sight of 1~Mpc. The units are $10^{-12}$~dyne~cm$^{-2}$. {\it{Bottom right: }}Map of the entropy derived from the best fit temperature and emission measure, assuming a length scale along our line of sight of 1~Mpc. The units are keV~cm$^{2}$. The white and black circles indicate the high surface brightness low entropy region and the region at the centre of the galaxy distribution, respectively. These we use for subsequent spectral analysis. The X-ray isophotes from the wavelet decomposed image are overplotted.} 
\label{fig:maps}
\end{figure*}

\begin{figure*}
\begin{minipage}{0.33\textwidth}
\includegraphics[width=0.98\textwidth,clip=t,angle=0.]{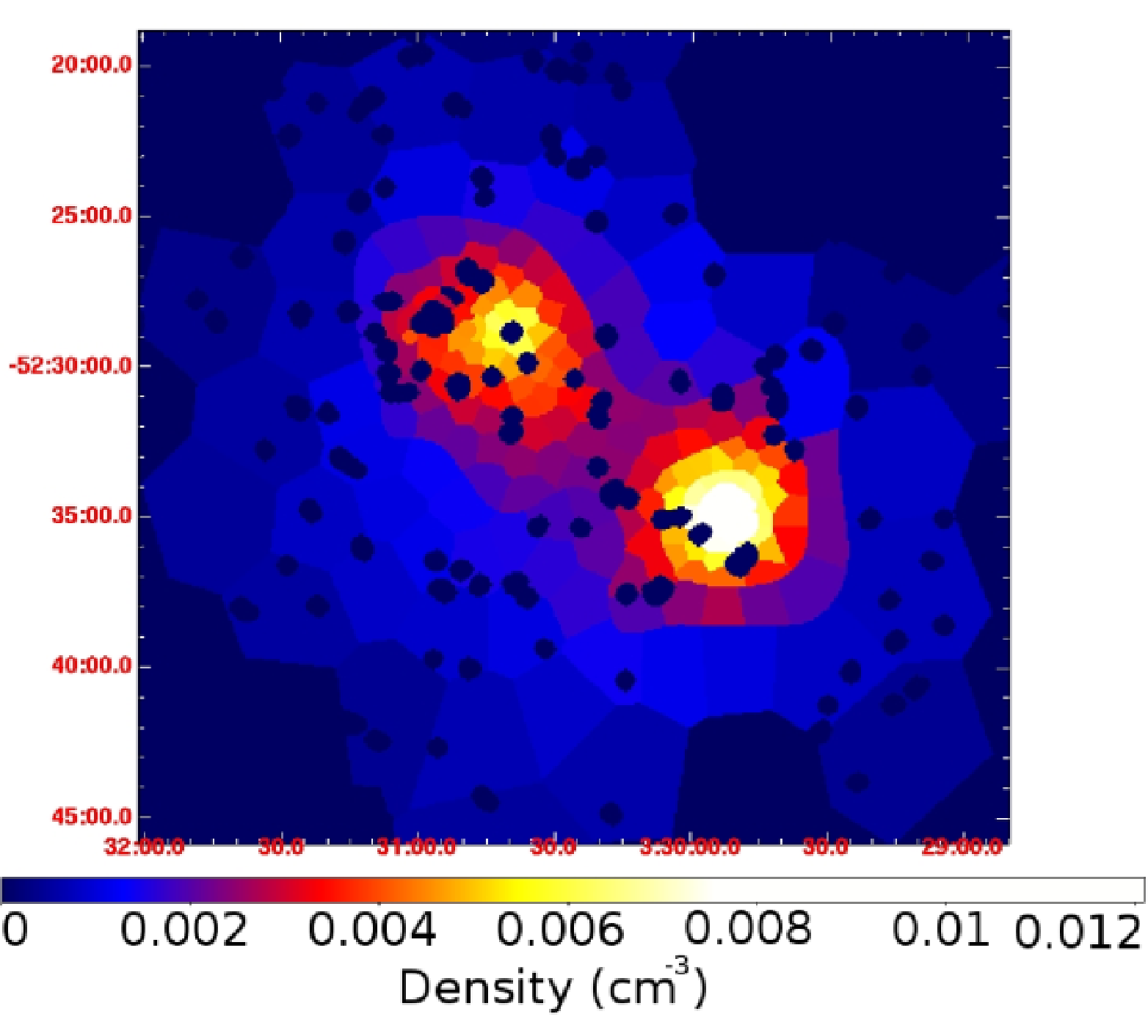}
\end{minipage}
\begin{minipage}{0.33\textwidth}
\includegraphics[width=0.98\textwidth,clip=t,angle=0.]{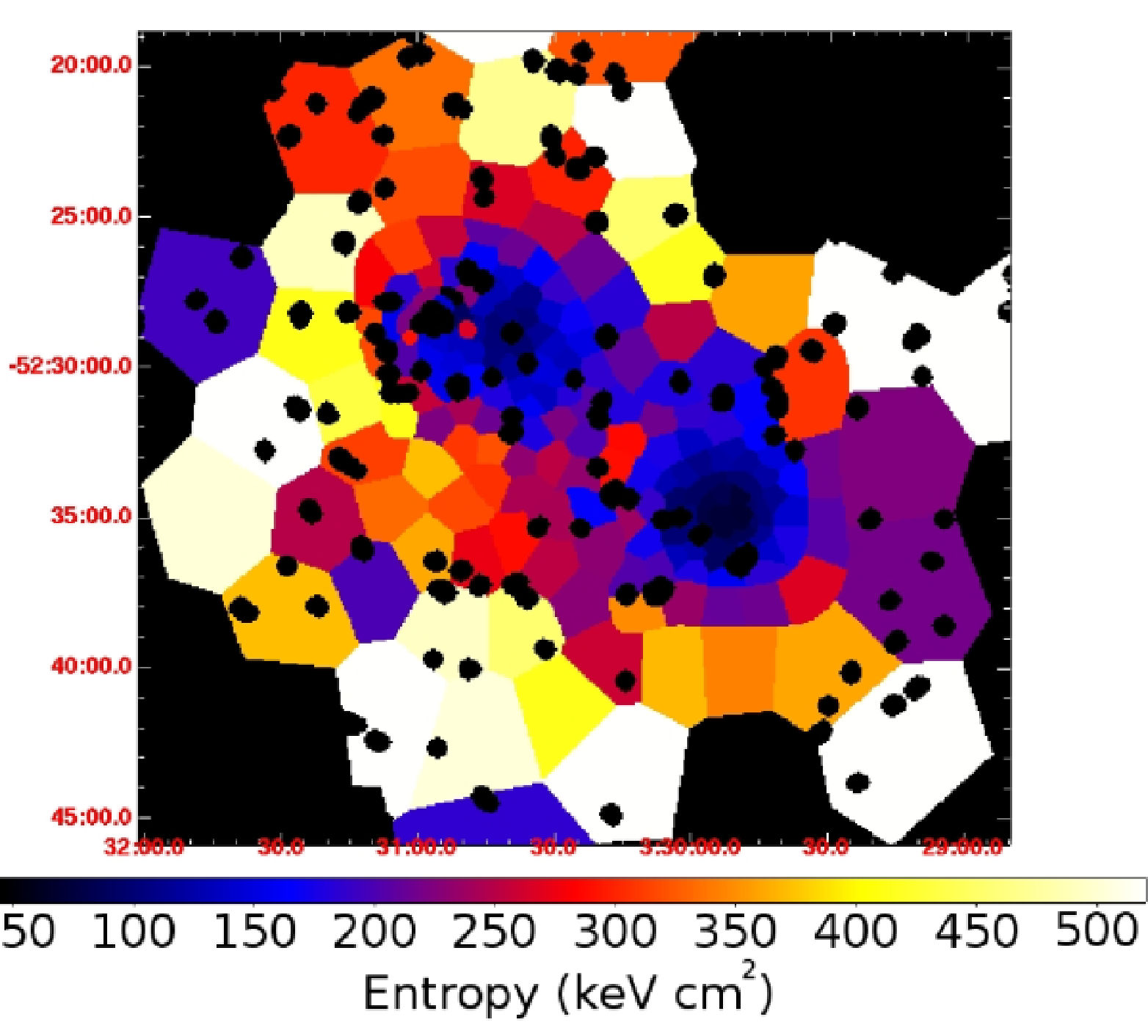}
\end{minipage}
\begin{minipage}{0.33\textwidth}
\includegraphics[width=0.98\textwidth,clip=t,angle=0.]{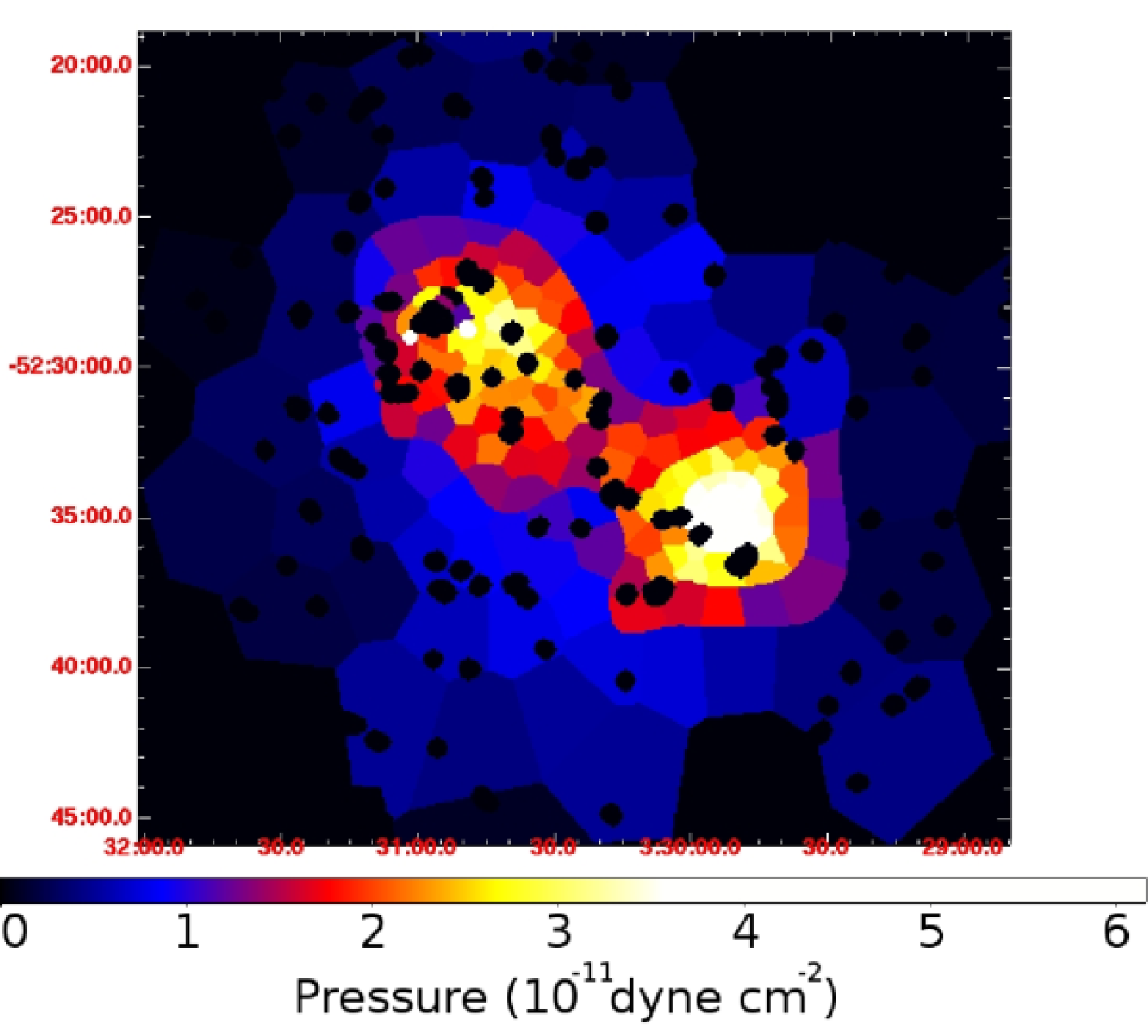}
\end{minipage}
\caption{{\it{Left panel: }}Density map of the ICM. {\it{Central panel: }} Map of the entropy of the ICM. {\it{Right panel: }} The pressure map of the ICM. The thermodynamic properties are calculated using volume estimates for the spectral extraction regions described in \citet{henry2004} and \citet{mahdavi2005}. The volumes are calculated assuming two centres, one at the SW X-ray peak, and one at the X-ray bright low entropy region at the possible centre of the gravitational potential of A3128. } 
\label{fig:mapsvol}
\end{figure*}

For making 2D maps of spectral parameters, we select regions according to the cluster surface brightness in the 3.5--7.5~keV band. Considering a optically thin plasma model \citet{forman2006} showed that for Chandra ACIS-I spectral response the flux $F$ in the $3.5$--$7.5$~keV band coming from a unit volume with a given pressure depends
only weakly on the gas temperature (over the 1--3~keV temperature range and for metallicity of $\sim$0.7 solar). For XMM-Newton EPIC and for the range of temperatures characteristic for A3128 (3--5 keV) this is also approximately correct: a factor $\eta$ relating the pressure and the square root of the flux
$\eta=(\epsilon(T)/T^2)^{1/2}$ decreases monotonically as a function of temperature by about 20\% (for metallicity of 0.45 solar). Here
$\epsilon(T)$ is the gas emissivity in the 3.5--7.5 keV band (with account for the spectral response of the instrument). Therefore, areas with similar surface brightness in the 3.5--7.5 keV band, provide us with contiguous regions without strong pressure discontinuities. We note here that our final maps of spectral parameters depend very weakly on
the procedure of region selection.

We then use the Voronoi tessellation method  \citep{cappellari2003,diehl2006} to further bin the selected areas with a stable minimum signal-to-noise ratio of $S/N=33$, which is needed for relatively accurate temperature determination.
We identify 213 independent regions from which we extract the spectra. For each region we compute a spectral redistribution file and an ancillary response file. We fit the spectrum of each bin individually with a single temperature thermal plasma model. The abundances of all elements except Fe in our model are fixed to 0.4 times the solar value. The emission measure, temperature, and Fe abundance are free parameters in the fit. From the best fit emission measure and temperature, we calculate for each extraction region the density $n$, entropy $s\equiv kT/n^{2/3}$, and the pressure $P\equiv nkT$, assuming a length scale along our line of sight of 1~Mpc. As a first step, we use a constant length scale in order to calculate the projected values of the thermodynamic properties without making any assumptions about the cluster centre and gas distribution.  

In order to subtract the emission of the background cluster from the 2D maps, we include in the fitted model for each bin with a mean distance smaller than 3.5\arcmin\ from the NE X-ray surface brightness peak, an additional thermal component. We fix its temperature, metallicity, and redshift to the mean values determined for the background cluster from the global fit: 5.14~keV and 0.47 solar, respectively (see Sect. \ref{globprop}). Using the best fit beta model to the background cluster (see Sect. \ref{secprofiles}) we calculate the normalization of the emission of the background cluster for each bin. By including this component in the fitted model, we effectively subtract its emission and determine the best fit parameters for the foreground cluster. However, since the core of the background cluster is not radially symmetric and it has a strongly elongated morphology, at radii smaller than 0.8\arcmin\ from the NE peak its emission cannot be properly subtracted. Therefore, we exclude the bins with a distance smaller than 0.8\arcmin\ from the NE peak from the analysis.

At the top left panel of Fig.~\ref{fig:maps} we show the map of the projected emission measure per square arcminute integrated along the line of sight, after the emission of the background cluster was subtracted. We see, that the surface brightness of the X-ray emission of A3128 has a peak at about 1.5\arcmin\ West-Southwest from the original NE peak. This region, as we show in the top right panel of Fig.~\ref{fig:maps} has a temperature between 3.0--3.5 keV. While the pressure is relatively high in this region (see the lower left panel of Fig.~\ref{fig:maps}), the entropy is low (lower right panel of Fig.~\ref{fig:maps}). This indicates that this region may be the centre of the gravitational potential of A3128. 
The centre of the galaxy distribution of A3128 is about 2.8\arcmin\ South of this region \citep{dalton1997,rose2002}.

In order to better compare the spectral properties of these two regions: the X-ray bright low entropy region and the region at the centre of the galaxy distribution, we extract a spectrum from a circular area with a radius of 1.26\arcmin\ centred on the new X-ray brightness peak ($\alpha=3^{\mathrm{h}}30^{\mathrm{m}}40^{\mathrm{s}}$, $\delta=-52^{\circ}28$\arcmin$50$\arcmin\arcmin) and a spectrum from a circular area with a radius of 1.5\arcmin\ centred on the optical centre ($\alpha=3^{\mathrm{h}}30^{\mathrm{m}}43^{\mathrm{s}}$, $\delta=-52^{\circ}31$\arcmin$30$\arcmin\arcmin) reported by \citet{rose2002}.
The spectral extraction regions are indicated by circles on the maps in Fig.~\ref{fig:maps}. The best-fit values of the single temperature thermal fits to the spectra are shown in Table~\ref{lowhighent}. We also show in the table the entropy and pressure values calculated for the volume of a sphere with a radius corresponding to the radius of the extraction region. The results confirm that the high surface brightness region has a factor of $\sim$1.5 lower entropy than the area at the optical centre of the galaxy distribution. The pressure at the high surface brightness region is a factor of $\sim$1.3 higher than the pressure at the centre of the galaxy distribution. The fit results indicate that the Fe abundance of the X-ray bright region is lower than the Fe abundance at the centre of the galaxy distribution. Under the assumption of hydrostatic equilibrium, the pressure peak combined with low entropy are unambiguous signs of the centre of the gravitational potential. 

The X-ray bright region at the possible centre of the gravitational potential of A3128 in the NE is separated from the bright SW core by a surface brightness depression. The pressure map indicates that the dark matter potential well associated with the SW core is connected with the dark matter potential in the NE. This is possibly the region where the dark matter potentials of the SW core and of A3128 overlap. We note that the ``bridge'' between the SW and NE in the pressure map coincides with an apparent chain of galaxies seen in the Digitalised Sky Survey image. Unfortunately, non of these galaxies has a known radial velocity.

In Fig.~\ref{fig:mapsvol}, we show density, entropy, and pressure maps determined assuming the X-ray emitting gas is associated with two gravitational potentials. The thermodynamic properties are calculated using volume estimates for the spectral extraction regions described in \citet{henry2004} and \citet{mahdavi2005}. The volumes are calculated assuming two centres, one at the SW X-ray peak, and one at the X-ray bright low entropy region at the possible centre of the gravitational potential of A3128. For each polygon the closest of the two centres is assumed to be the centre in the volume calculation. 

\section{Discussion}
\label{discussion}

\subsection{The background cluster associated with the NE X-ray peak}

While at the expected energy we do not detect the strong Fe~K line in the spectrum extracted from the NE X-ray peak, we detect line emission at the energy corresponding to the line energy of Fe~K emission redshifted by $z=0.44$. Therefore we conclude that the NE X-ray peak observed toward A3128 is not associated with the surviving ICM of a group falling in supersonically to the cluster as previously thought \citep{rose2002}, but it is a distant luminous cluster of galaxies at redshift $z=0.44$. Subsequent optical spectroscopic observation of the distant radio bright galaxy in the centre of the NE X-ray peak (radio source SUMSS~J033057-522811) with the Magellan telescope revealed a redshift of $z=0.44$ confirming its association with the cluster seen in X-rays. 

The red magnitude of the galaxy $m_{\mathrm{R}}=18.5$ is typical for cD galaxies at this redshift \citep[e.g.][]{vikhlinin1998}. The Magellan observation also confirmed the gravitational arc around the galaxy, the presence of which was previously suggested by the images obtained with the 0.9~m CTIO telescope \citep{rose2002}. We estimate the enclosed total mass within the radius of the lens to be $<5\times10^{12}$~M$_{\odot}$, which is within a factor of 2 consistent with the total mass of M~87 at a similar radius. The optical data thus strongly suggest that SUMSS~J033057-522811 with properties characteristic for cD galaxies is the dominant galaxy of the cluster in the background. 

The X-ray morphology of the background cluster is strongly elongated showing that its core
is not relaxed and that it is possibly undergoing a merger. However,
its observed X-ray luminosity of $L=6.9\times10^{44}$~erg~s$^{-1}$, temperature
k$T=5.14$~keV, and derived mass $M=3.4\times10^{14}$~M$_{\odot}$ agree well with
the generally observed $L_{X}$--$T$ and $M$--$T$ scaling relations
\citep[e.g.][]{wu1999,pratt2006}. 
Further deep optical observations are needed to characterise the properties of the background cluster.

\subsection{The nature of the SW surface brightness peak}

The SW surface brightness peak is centred on an apparent compact group of galaxies. Compared to other clusters of galaxies with similar temperatures, its core radius is small. Such small core radii are observed only in clusters with cooling cores, or in clusters undergoing merging events where the cluster profile is truncated. While the radial Fe abundance distribution has a strong peak at the centre of the group, the temperature distribution is flat, with no indication for a cool core. The cooling time in the centre of the SW core, assuming the gas cools isobarically, is $4.5\times10^{9}$~yr. Its estimated total mass of $M_{500}=1.1\times10^{14}$ is a factor of 2 lower than the expected mass of a 3.4~keV cluster \citep[e.g.][]{pratt2006}. But since the value of $\beta=0.3$ was determined by fitting the surface brightness profile at radii much smaller than $r_{500}$, the estimated $M_{500}$ is highly uncertain. 
The radial velocity of the dominant galaxy of this apparent group (ENACS~75) is $V=19252$~km~s$^{-2}$ \citep{katgert1998}, which is higher by $\approx1500$~km~s$^{-1}$ than the mean radial velocity of galaxies in A3128. Unfortunately, we know the radial velocity for only one more galaxy of the apparent group (ENACS~78), which is lower than that of the dominant galaxy: $V=18380$~km~s$^{-1}$ \citep{katgert1998}, but still higher by $\approx500$~km~s$^{-1}$ than the mean radial velocity of the cluster. 
The sound speed corresponding to the mean cluster temperature of $\sim$3.5~keV is 960~km~s$^{-1}$, which means that if the SW peak is the remaining core of a group or a cluster merging with A3128 at a relative velocity of $\approx1500$~km~s$^{-1}$, then the merger is supersonic. 

On the smaller spatial scales, the X-ray emission of the SW peak appears to follow the galaxy mass distribution as it is slightly displaced toward the West and clearly peaks on the galaxy ENACS~75. However, the association of the galaxy ENACS~75 with the SW X-ray peak in radial velocity is difficult.
The redshift of the SW peak determined from the energy centroid of the Fe~K line corresponds to a radial velocity of $V=18000\pm300$~km~s$^{-1}$, and the redshift of the ``tail'' region, which we might consider as the diffuse emission of A3128 corresponds to a radial velocity of $V=17400\pm900$~km~s$^{-1}$. However, the systematic uncertainties on these values are larger than the quoted statistical errors. The EPIC detectors are known to have gain problems which are both time and position dependent. The absolute error in the redshift determination can be as large as 1500~km~s$^{-1}$ (Simionescu et al. in prep.). Therefore, the radial velocity of the hot gas is within the systematic uncertainties consistent with the radial velocity of the galaxies.

If the SW peak is a core of a group or of a cluster that moves through the ICM of A3128, than its gas is being stripped by ram pressure. The gas of the infalling cluster is stripped at radii where the thermal pressure of the hot gas in the infalling group is too small to balance the sum of the thermal and ram pressure of the cluster ICM. This happens at radii larger than the radius where the thermal pressure of the infalling cluster is equal to the thermal pressure at the stagnation point \citep[e.g.][]{markevitch2007}. The ratio of thermal pressures at the stagnation point $p_0$ and in the free stream, $p_1$, for Mach numbers relative to the sound speed in the free stream region $M>1$ is \citep{landau1959}:
\begin{eqnarray}
\frac{p_{0}}{p_{1}}=\left( \frac{\gamma+1}{2}\right)^{\frac{\gamma+1}{\gamma-1}}M^{2}\left( \gamma-\frac{\gamma-1}{2M^{2}}\right)^{-\frac{1}{\gamma-1}},
\end{eqnarray} 
where $\gamma=5/3$ is the adiabatic index of the gas. For the pressure in the free stream region, $p_1$, we assume the value determined assuming a density of $n=1\times10^{-3}$~cm$^{-3}$ and a temperature of 3.5~keV. For the pressure of the SW core at the given radius, we assume a density profile of $n(r)=n_{0}(1+(r/r_{c})^{2})^{-\frac{3}{2}\beta}$, where $n_{0}=1.2\times10^{-2}$~cm$^{-3}$, $r_{c}=30$~kpc, and $\beta=0.3$. We find that at a Mach number of $M=1.5$ (inferred from the radial velocity difference of $\sim1500$~km~s$^{-1}$ between ENACS~75 and A3128, and from the sound speed $v_{\mathrm{s}}=960$~km~s$^{-1}$ in the ICM), the gas outside the radius of $r=90$~kpc should be stripped, and only the ICM of the merging core inside of this radius should be surviving. 

Such a merger would produce a tail of low entropy and high Fe abundance.
However, in the images and in the maps of the thermodynamic properties we do not observe any obvious signs of a merger for the SW peak.  
Moreover, we verified that the Fe abundance distribution is consistent with being symmetric and is not enhanced in any direction. These observed properties might indicate that the merger is occurring just along our line of sight. The merger is compressing and heating the ICM, which might explain the observed lack of a cool core. 
A shock with a Mach number $M=1.5$ heats the gas by a factor of 1.5, which means that for the cluster temperature of 3.5~keV the gas temperature at the stagnation point should be 5.25~keV. If we fit the spectrum extracted from the circular region with a radius of 1.5\arcmin\ centred on the SW core with two thermal models, with the temperature of one thermal component fixed to k$T=5.25$~keV, the fit improves compared to the single temperature model (see Sect.~\ref{globprop}).
The reduced $\chi^2$ improves from 1.30 to 1.16. We obtain a best fit temperature of k$T=2.23\pm0.17$~keV for the cooler component, and best emission measures of $Y=(5.1\pm0.5)\times10^{65}$~cm$^{-3}$ and $Y=(4.7\pm0.5)\times10^{65}$~cm$^{-3}$ for the hotter and cooler gas, respectively. This indicates that much of the emission of the SW core might be due to the shocked gas from a merger happening along our line of sight. 
As the enriched stripped gas is trailing behind the core we are seeing it in projection which explains the enhanced Fe abundance within the projected distance of $\sim$120~kpc. The velocity difference between the two galaxies of the compact group (ENACS~75, ENACS~78) may also be explained by tidal stretching of the group along our line of sight during the passage of the group through the cluster. 

Alternatively, the SW core might be a group still falling toward A3128, which we observe projected on the ICM of the cluster. In this case the ICM of the group still did not start to interact with the ICM of the cluster, and ENACS~78 and 75 are only projected close to each other by chance. However, this scenario does not explain the lack of the temperature gradient and of a cool core in the group.

\subsection{The diffuse ICM}

After subtracting the emission of the $z=0.44$ cluster, we identified a new region with an enhanced surface brightness to the West-Southwest of the centre of the background cluster. This region has also an enhanced pressure and low entropy. It is at a distance of $\sim$2.8\arcmin\ from the centre of the galaxy distribution given by \citet{rose2002}. However, the entropy of the ICM at the centre of the galaxy distribution is a factor of 1.4 higher and the pressure is lower than that of the region with the enhanced X-ray surface brightness. Because of the great dynamical complexity of the system, with infalling groups and filaments identified in position-position and position-redshift diagrams, the determination of the cluster centre from the galaxy distribution is highly uncertain. Based on the thermodynamic properties of the ICM, we conclude that the enhanced surface brighness region, centred at $\alpha=3^{\mathrm{h}}30^{\mathrm{m}}40^{\mathrm{s}}$, $\delta=-52^{\circ}28$\arcmin$50$\arcmin\arcmin, is the centre of the gravitational potential of the cluster A3128. The position of this region is also more consistent with the centre of the extended low surface brightness X-ray emission.

The images and the maps of thermodynamic properties do not reveal any obvious shocks. We only see two candidates, one to the Northeast of the background cluster and one to the Southwest of SW core. Both shock candidates exhibit temperature drops by $\sim$1~keV associated with a drop in entropy, and a factor of 2 drop in pressure. 
However, the surface brightness in these two areas is low and we cannot confirm the presence of surface brightness discontinuities in the regions where the maps indicate the presence of the shocks.  

As proposed by \citet{caldwell1997}, A3128 may have had encountered a merger with A3125 in the past, and as discussed by \citet{rose2002} the optical redshift data reveal a number of groups, some of them tidally distended into filaments after a close passage through A3128. 
The X-ray data combined with the optical redshifts suggest an ongoing merger with a group. Moreover, the inconsistency between the cluster centre determined based on the distribution of galaxies and based on the thermodynamic properties of the ICM, and the lack of bright galaxies at the newly identified centre of gravitational potential, further highlight the view that A3128 is a dynamically young, unrelaxed system. The unrelaxed nature of A3128 can be attributed to its location in the high density environment of the Horologium-Reticulum supercluster.

\section{Conclusions}

We have analyzed new deep XMM-Newton EPIC data of the cluster of galaxies Abell~3128 located in the Horologium-Reticulum supercluster. We found that:

\begin{itemize}
\item
The Northeast X-ray peak observed toward A3128 is a distant luminous cluster of galaxies at redshift $z=0.44$.
\item
The properties of the distant radio bright galaxy in the centre of the NE X-ray peak indicate that it is the cD galaxy of the cluster in the background. We detect a gravitational arc around the galaxy.
\item
The properties of the Southwest X-ray peak suggest that it is the core of a group merging with A3128 along our line of sight.
\item
Based on 2D maps of thermodynamic properties of the ICM determined after subtracting a model for the background cluster, we conclude that the enhanced surface brightness region at a distance of $\sim$2.8\arcmin\ from the centre of the galaxy distribution is the centre of the gravitational potential of the cluster A3128. The inconsistency between the cluster centres determined based on the distribution of galaxies and based on the thermodynamic properties of the ICM further highlights the view that A3128 is a dynamically young, unrelaxed system.
\end{itemize}

\begin{acknowledgements}
NW acknowledges the support by the Marie Curie EARA Early Stage Training visiting fellowship.
AF acknowledges a support from the NASA grant NNX06AF07G.
This work is based on observations obtained with XMM-Newton, an ESA science mission with instruments
and contributions directly funded by ESA member states and the USA (NASA). The Netherlands Institute
for Space Research (SRON) is supported financially by NWO, the Netherlands Organization for Scientific Research. 
\end{acknowledgements}

\bibliographystyle{aa}
\bibliography{clusters}

\end{document}